# Accurate and Robust Genomic Prediction of Celiac Disease using Statistical Learning

Gad Abraham[1,2], Jason A. Tye-Din[3,4,5], Oneil G. Bhalala[1], Adam Kowalczyk[2], Justin Zobel[2], and Michael Inouye[1,*]

[1]Medical Systems Biology, Department of Pathology and Department of Microbiology & Immunology, The University of Melbourne, Parkville 3010, Victoria, Australia
[2]NICTA Victoria Research Lab, Department of Computing and Information Systems, The University of Melbourne, Parkville 3010, Victoria, Australia
[3]The Walter and Eliza Hall Institute of Medical Research, 1G Royal Parade, Parkville 3052, Victoria, Australia
[4]Department of Medical Biology, The University of Melbourne, Parkville 3010, Victoria, Australia
[5]Department of Gastroenterology, The Royal Melbourne Hospital, Grattan St., Parkville 3050, Victoria, Australia.

* Correspondence should be addressed to Michael Inouye – minouye@unimelb.edu.au

## Abstract

Practical application of genomic-based risk stratification to clinical diagnosis is appealing yet performance varies widely depending on the disease and genomic risk score (GRS) method. Celiac disease (CD), a common immune-mediated illness, is strongly genetically determined and requires specific HLA haplotypes. HLA testing can exclude diagnosis but has low specificity, providing little information suitable for clinical risk stratification. Using six European CD cohorts, we provide a proof-of-concept that statistical learning approaches which simultaneously model all SNPs can generate robust and highly accurate predictive models based on genome-wide SNP profiles. The high predictive capacity replicated both in cross-validation within each cohort (AUC of 0.87—0.89) and in independent replication across cohorts (AUC of 0.86—0.9), despite differences in ethnicity. The models explained 30—35% of disease variance and up to ~43% of heritability. The GRS's utility was assessed in different clinically relevant settings. Comparable to HLA typing, the GRS can be used to identify individuals without CD with ≥99.6% negative predictive value however, unlike HLA typing, patients can also be stratified into categories of higher-risk for CD who would benefit from more invasive and costly definitive testing. The GRS is flexible and its performance can be adapted to the clinical situation by adjusting the threshold cut-off. Despite explaining a minority of disease heritability, our findings indicate a predictive GRS provides clinically relevant information to improve upon current diagnostic pathways for CD, and support further studies evaluating the clinical utility of this approach in CD and other complex diseases.

Supplementary Methods and Results are available at http://dx.doi.org/10.6084/m9.figshare.154193

# AUTHOR SUMMARY

Celiac disease (CD) is a common immune-mediated illness, affecting approximately 1% of the population in Western countries but the diagnostic process remains sub-optimal. The development of CD is strongly dependent on specific human leukocyte antigen (HLA) genes and HLA testing to identify CD susceptibility is now commonly undertaken in clinical practice. The clinical utility of HLA typing is to exclude CD when the CD susceptibility HLA types are absent, but notably, most people who possess HLA types imparting susceptibility for CD never develop CD. Therefore, while genetic testing in CD can overcome several limitations of the current diagnostic tools, the utility of HLA typing to identify those individuals at increased-risk of CD is limited. Using large datasets assaying single nucleotide polymorphisms (SNPs), we have developed genomic risk scores (GRS) based on multiple SNPs that can more accurately predict CD risk across several populations in "real world" clinical settings. Our GRS can generate predictions which optimize CD risk stratification and diagnosis, potentially reducing the number of unnecessary follow-up investigations. The medical and economic impact of improving CD diagnosis is likely to be significant, and supports further studies into the role of personalized GRSs for other genetically-linked human diseases.

# INTRODUCTION

Improving the diagnosis of celiac disease (CD), a common immune-mediated illness caused by dietary gluten, remains a clinical challenge [1,2]. Despite a prevalence of approximately 1% in most Western countries, lack of awareness and failure to implement appropriate serological, histological, and genetic testing means that less than 30—40% of those affected by CD are diagnosed [1,3-5]. Undiagnosed CD is associated with reduced quality of life, substantial morbidity, and increased mortality, however, prompt diagnosis and treatment lowers the burden of disease and may reduce the rate of complications such as osteoporosis, autoimmune disease, and malignancy. Optimizing the diagnosis of CD is now recognized as an important goal for clinicians [6].

CD is characterized by a variable combination of gluten-dependent clinical manifestations, CD-specific antibodies, and small bowel inflammation (villous atrophy) [7]. Traditional guidelines for the diagnosis of CD rely on demonstrating villous atrophy and improvement of symptoms, laboratory abnormalities, and/or small bowel inflammation upon exclusion of dietary gluten [8]. Current clinical practice is to screen for CD by detecting CD-specific serum antibodies, and then confirm the diagnosis by undertaking small bowel biopsy to demonstrate typical villous atrophy. Serologic screening for CD with transglutaminase-IgA antibodies is reported to be highly sensitive and specific for CD (both >90%), imparting a high positive predictive value (PPV) of over 90% when assessing most populations [9,10], although the PPV can fall to 45—70% in community screening settings [11,12]. In practice, serological and histological assessments have technical limitations that generate both false negative and false positive diagnoses.

A key feature of CD is its strong dependence on the presence of susceptibility genes encoding for HLA DQ2.5, DQ8, and/or half the HLA DQ2.5 heterodimer (typically DQ2.2), seen in approximately 99.6% of all patients with CD [13]. These genes encode immune-recognition molecules which facilitate CD4+ T cell recognition of specific gluten-derived peptides, a critical step in disease pathogenesis [14-18]. Recognizing the crucial role of these genes, the latest consensus diagnostic guidelines for CD recommend testing for these HLA heterodimers (HLA typing) as a first-line investigation for asymptomatic individuals identified at-risk of CD, such as 1st-degree relatives of an affected individual or those with suggestive symptoms [7]. However, a major flaw of HLA typing as a diagnostic tool is that a substantial proportion, typically reported as 30—40%, of the community express HLA DQ2.5, DQ8, and/or DQ2.2, making the presence of these HLA types poorly predictive and of low specificity for CD [13]. Indeed, a recent Australian population study revealed that 56% of the community possessed at least one of these CD susceptibility haplotypes [5]. Thus while HLA typing can exclude CD in the community with high confidence when the susceptibility haplotypes are absent, these haplotypes will be present in 30—56% of the population, the majority of whom would not have CD. Therefore, if assessed as a stand-alone test, HLA typing has exceptionally high sensitivity and negative predictive value (NPV), but very poor specificity and low positive predictive value (PPV) for CD. Since a positive result poorly predicts the presence of CD, HLA typing is not useful as a stand-alone diagnostic tool for CD. While the relative-risk for CD can be stratified based on the HLA subtype (CD risk DQ2.5 > DQ8 > DQ2.2) [19], these categories have low positive predictive value and do not provide clinically-informative attribution of CD risk [20]; HLA results are therefore interpreted as a binary outcome: CD susceptibility positive or negative. Despite these limitations, HLA typing is now widely utilized in clinical practice and typically determined using polymerase chain-reaction sequence specific oligonucleotide (PCR-SSO) hybridization, which is time and labor intensive, and costly (AU $120/sample, Medicare; in the USA cost varies, but is typically US $150/sample or greater).

It is important to distinguish between three different approaches to analyzing the HLA region for association with CD. The first approach, currently in clinical practice, is HLA typing, as described above, where the HLA result is considered a binary variable and its utility is to exclude CD. A second approach, such as that taken by Romanos et al., utilizes the same HLA-DQ haplotypes, stratifies individuals into several nominal risk levels then fits a statistical model to empirically estimate the true risk in each group [21,22]. While HLA-DQ haplotypes may be inferred from typing several HLA SNPs, importantly, the HLA SNPs are only used to assign the HLA type and the SNPs themselves are not directly modeled. The third approach, such as that used here, is based on direct concurrent modeling of many thousands of individual SNPs for association with CD, to produce a more fine-grained predictive "genomic risk score" (GRS).

GRSs have been enabled by the advent of genome-wide association studies (GWAS), which perform unbiased testing of many thousands of SNPs for association with CD. Using GWAS, recent studies have identified multiple non-HLA SNP associations with CD [23,24]. GWAS are primarily concerned with the detection of variants associated with disease in order to gain insight into the disease etiology and genetic architecture. Due to the high number of significance tests, controlling for false positive associations is a major concern. Therefore, SNP-based risk scores have tended to be constructed from the SNPs found to be significantly associated with the disease status [22,25]. However, due to the stringent multiple-testing corrections performed in GWAS there may be other SNPs that fail to achieve genome-wide significance but may be predictive of disease status nonetheless and including them in the model could potentially result in higher predictive ability than achievable by models based solely on genome-wide significant SNPs. In contrast to the GWAS approach, the main overriding aim of a diagnostic GRS from a clinical perspective is to achieve maximal predictive capacity, and the inference of genetic architecture is secondary.

We have recently designed computational algorithms which efficiently fit L1-penalized multivariable classification models to genome-wide and whole-genome SNP data [26]. Such models were then shown to be preferable to several other methods such as the standard method of summing the per-SNP log odds (polygenic score) [27], mixed effects linear modeling [28,29], and unpenalized logistic regression, with both better precision for detecting causal SNPs in simulation and better case/control predictive power [30]. These advantages were consistent across several complex diseases, including two cohorts of CD (UK1 and UK2). However, the diagnostic implications of penalized models have not been previously examined nor has the robustness of such models in other populations or the advantage over HLA-typing approaches. In contrast to existing studies that examine a small number of genome-wide significant SNPs, we have shown that many more SNPs (tens to hundreds) are required to achieve optimal predictive ability for CD. Further, the standard GWAS approach of considering each SNP separately when estimating its effect size does not consider its correlation with other SNPs. We have shown that unpenalized predictive models based on these top SNPs suffer from lower predictive ability than L1-penalized models since the pre-screening introduces multiple highly-correlated SNPs into the model, of which a substantial proportion may be redundant in terms of contribution to the predictive ability. Similar L1-penalized approaches have also recently been successfully applied to inflammatory bowel disease case/control Immunochip data, where models based on several hundred SNPs have led to high predictive ability [31].

Here, we provide a proof-of-concept that the GRS for CD, induced by L1-penalized support-vector machine models, is able to achieve a predictive capacity and robustness that provides information not afforded by current diagnostic pathways utilizing HLA typing alone. This GRS has the potential to provide greater clinical diagnostic utility by enabling each individual to be assigned a more informative risk score beyond the simple designation of "CD susceptible" or "CD non-susceptible", or even "high risk" versus "low risk". To enable useful comparisons between diagnostic approaches, we model the GRS as a stand-alone test to "diagnose" CD, while at the same time acknowledging that real world clinical practice will need to draw upon clinical history, CD-specific serology, and small bowel histology to confirm the diagnosis of CD. We assess the predictive power of the GRS both in cross-validation and in independent validation, across six different European cohorts, showing that the models strongly replicate. We test our GRS on three other autoimmune diseases: type 1 diabetes (T1D), Crohn's disease, and rheumatoid arthritis, finding some predictive ability for T1D status but none for the others, thus largely supporting the specificity of the scores for CD. To overcome limitations of previous studies utilizing GWAS case/control studies, where ascertainment bias incurs substantially higher rates of false positive results, we undertake genomic prediction of CD in "real world" settings where the prevalence of CD is far lower, and evaluate the performance of the GRS by assessing the PPV and NPV at several levels of CD prevalence. Unlike HLA typing, the GRS allows flexibility in determining who is considered at higher risk for CD by selecting a clinically-determined user-specified threshold. We demonstrate how these scores can be practically applied at various prevalence levels to optimize sensitivity and precision. Finally, we show how the model can be calibrated to produce accurate predicted probabilities of disease.

## RESULTS

An overview of our analysis workflow is shown in Figure 1. We analyzed five CD datasets on the Illumina array platform: *UK1* and *UK2* (British descent), *Finn* (Finnish descent), *IT* (Italian descent), and *NL* (Dutch descent). We also utilized a dataset run on a fine-mapping array: *IMM* (Immunochip of British descent) (see Methods and Table 1). We have previously analyzed the AUC achievable in the UK1 and UK2 datasets [30].

We trained L1-penalized support vector machines (SVM) [26] on the genotype data, including all post quality control (QC) autosomal SNPs unless otherwise indicated. These models are sparse models, (Methods), and varying the penalty induces models based on different number of SNPs with non-zero coefficients, and we investigated the performance for various degrees of sparsity. The models were fitted to all SNPs across the genome simultaneously. For each model's induced risk score, we estimated the area under the receiver operating characteristic curve (AUC) and the explained phenotypic variance [32], which depends on the assumed prevalence of disease.

## CROSS-VALIDATION IN EACH DATASET

We used 10x10 fold cross-validation to estimate the AUC and explained phenotypic variance (on the 0-1 liability scale) within each dataset. The explained phenotypic variance was derived from the AUC assuming a population prevalence of $K=1\%$. All cohorts showed high AUC in cross-validation (Figure 2a), with the Finnish and Italian cohorts having a maximum of 0.89, followed by the UK1 cohort (AUC=0.88), and finally the UK2 and Dutch cohorts with a maximum AUC of 0.87. Both the UK1 and the Italian cohorts peaked at ~64 SNPs with non-zero weights, whereas the rest peaked at ~250 SNPs. Subsampling of the individuals in the UK2 dataset indicated approximately diminishing returns with 80% of the sample size having the same AUC as 100% (Supplementary Figure 1). Consistent with this, combining the UK1 and UK2 datasets did not increase AUC beyond UK2 alone (results not shown). We also note that some of the control samples were population-based and were not explicitly screened for celiac disease, thus, ~1% may be cryptic CD cases which potentially underestimates prediction performance in downstream analyses. These AUCs correspond to explained phenotypic variance of 30—35% (Figure 2b). Assuming a CD heritability of 80%, this translated to an explained genetic variance of 37—43%.

## EXTERNAL VALIDATION BETWEEN DATASETS

While cross-validation provides an estimate of the model's ability to generalize to unseen datasets, choosing the model with best AUC may lead to so-called "optimization bias" (also called "winner's curse") [33,34], potentially manifesting as lower performance in independent validation. Additionally, cross-validation cannot compensate for any intra-dataset batch effects, as these would be present in both the training and testing folds, potentially artificially inflating the apparent predictive ability. To assess whether the models suffered from optimization bias and to exclude the possibility of strong intra-dataset batch effects, we performed external validation: based on the results of the cross-validation, we selected the best models trained on the UK2 dataset, tested them on the UK1, Finn, IT, and NL datasets, without any further tuning, and computed the receiver-operating characteristic (ROC) curves (Figure 3a). Overall, the models trained on the UK2 cohort showed high reproducibility on the other cohorts, achieving high AUC of 0.89—0.9 in the Finnish and UK1 datasets, indicating negligible optimization bias from the cross-validation procedure. We also examined the replication of different SNP sets (all autosomal, MHC, and non-MHC) trained on the UK2 dataset and tested on the others (Supplementary Figure 2). The trends observed in cross-validation, namely, similar performance for MHC and all autosomal SNPs, with lower but still substantial performance of non-MHC SNPs, was observed in all external validation experiments.

## COMPARISON OF GENOMIC RISK SCORE WITH METHODS BASED ON HLA TYPING

Since HLA typing is commonly used for assessing CD risk status, we sought to compare the performance of an approach based on inferred HLA types with the GRS. We utilized the approach of Romanos et al. [21] on the Immunochip data, which relies on both HLA types and 57 non-HLA Immunochip SNPs (including one chrX SNP). Since directly measured HLA types were not available for our datasets, we imputed *HLA-DQA1* and *HLA-DQB1* haplotype alleles using HIBAG [35] and derived the presence of DQ2.2 / DQ2.5-homozygous / DQ2.5-heterozygous / DQ8 heterodimer status. The coefficients for the HLA risk types in the HLA+57 SNP model were not available for the Romanos et al. method, and we estimated these from our data. For application of the GRS method, we trained models on 18,252 autosomal SNPs from UK2 (the subset shared between the Illumina 670 and Immunochip) then externally validated these models on the Immunochip data. We trained three separate models: All autosomal SNPs, MHC SNPs, and autosomal non-MHC SNPs. As seen in Figure 3b, the GRS trained on either all SNPs or the MHC SNPs yielded higher AUC (0.87) than the Romanos HLA+57 SNPs (AUC=0.85) or HLA type alone (AUC=0.8). The predictive power of the GRS induced by SNPs outside the MHC was lower, but still substantial, at AUC=0.72. We also performed similar analyses on the rest of the datasets (UK2 in cross-validation, then externally validated on UK1, Finn, NL, IT, and UK1), comparing the GRS with the HLA type and with analysis of HLA tag SNPs [36] commonly used to infer HLA types since the 57 non-HLA SNPs used by Romanos were not available on these platforms. As shown in Supplementary Figure 2, the HLA type approach had consistently lower AUC (0.795—0.86) than analysis of the individual HLA tag-SNPs of Monsuur (AUC of 0.85—0.876, modeled using logistic regression on the SNPs in the UK2 dataset, and tested on the other datasets) and substantially lower than the GRS (AUC of 0.86—0.894).

Overall, these results show that the L1-penalized SVM approach which modeled the SNPs directly was able to extract more information from the HLA region than the coarse-grained HLA haplotype model, either with or without the addition of the 57 non-HLA SNPs, corresponding to a gain in explained phenotypic variance of 3.5% over the best Romanos et al model in the Immunochip data.

### SPECIFICITY OF THE GENOMIC RISK SCORE

We investigated whether the models of CD were predictive of case/control status in other immune-mediated diseases, specifically type 1 diabetes (T1D), rheumatoid arthritis (RA), and Crohn's Disease/Inflammatory Bowel Disease (Crohn's) from the WTCCC [37]. We took the SNPs that appeared on both the UK2 Illumina and WTCCC Affymetrix 500K arrays (after QC), resulting in 76,847 autosomal SNPs. Despite the substantial reduction in the number of SNPs from the original data, we observed only small reductions in AUC in the restricted UK2 dataset in cross-validation, indicating that most of the predictive information was retained in the reduced SNP set (AUC=0.85 at ~200 SNPs). The models trained on the UK2 were subsequently tested on the T1D, RA, and Crohn's datasets. We also used the Finnish CD dataset as external validation to ensure that the high predictive performance observed in cross-validation on UK2 was replicated on other CD datasets and not degraded by using fewer SNPs. Overall, the models showed some predictive ability of T1D (AUC=0.69), consistent with previous findings showing shared genetics between T1D and CD [38,39] (see Supplementary Figure 3 for results for the MHC and non-MHC SNPs in T1D), and had very low performance (AUC 0.51—0.54) on the RA and Crohn's datasets. In contrast, performance on the Finnish CD cohort was only slightly lower (AUC=0.85) compared with the full SNP set, again confirming that the CD models replicated across ethnic cohorts despite using a reduced set of SNPs (Figure 3c).

### ANALYSIS OF A COMBINED DATASET

All CD datasets showed consistently high AUC both in cross-validation and in external validation, indicating that the risk of substantial confounding of the case/control status by ethnic cohort, that is, population stratification or strong intra-cohort batch effects, was low. Therefore, in order to increase statistical power in comparing the performance of the models, we created a combined dataset consisting of the Finnish, Dutch, and Italian cohorts, totaling 5165 samples (1947 cases and 3218 controls, 512,634 SNPs). This combined dataset may be more representative of a real screening scenario, where individuals of different ethnicities are being screened for CD. Figure 4a shows kernel density estimates of the predicted risk scores for cases and controls in the combined dataset, where the scores are based on models trained on the UK2 dataset as previously described. As expected from the high AUC, there was substantial separation between the score distributions for the two classes. Also shown is the percentage of the combined population corresponding to a range of GRS thresholds (Figure 4b).

### POSITIVE AND NEGATIVE PREDICTIVE VALUES UNDER DIFFERENT PREVALENCE SETTINGS

The prevalence of CD in the general population (for the purpose of this modeling taken to be 1%) is much lower than the prevalence in the case/control datasets, where the cases are substantially over-represented owing to the study design, causing problematic ascertainment bias. Considering the prevalence as the prior probability of a person having the disease (without knowing their genetic profile), then unless the likelihood of disease given the genotype is high as well, the posterior probability of disease will remain low. To quantify the predictive performance of our models while accounting for the prevalence, we estimated the precision of our models trained on UK2 on the Finn+NL+IT combined dataset. We down-sampled the cases in the combined dataset to simulate settings with different CD prevalence levels (1%, 3%, 10%, and 20%), and estimated precision and sensitivity in the test data, repeated in 50 independent simulations for each prevalence level (Figure 5a). The precision here is equivalent to the PPV [40], as the precision is estimated in data with the same prevalence as assumed by the PPV. The PPV is the posterior probability of having the disease given a positive diagnosis and the NPV is the posterior probability of not having the disease given a negative diagnosis, where a perfect model would have PPV = NPV = 1. Note that the lowest NPV achievable is 1−prevalence, which translates to seemingly high NPV values in the low-prevalence setting, rendering NPV less useful for assessing classifiers in such settings, as even a weak classifier can achieve apparently high NPV.

Population screening for CD is not currently accepted practice. Most evidence supports an active case-finding strategy, where patients with risk factors for CD, and therefore higher pre-test probability of CD than the population-wide average, are identified by their primary practitioner and screened. For example, the prevalence of CD in patients with a first-degree relative with CD is 10% or higher [41,42], and the prevalence of CD in patients with T1D is 3—16% [43]. The increased CD prevalence in these groups of patients improves the diagnostic performance of the GRS. To examine the effect of prevalence on PPV, we first employed the GRS in a population-based setting (prevalence of 1%), which resulted in a PPV of ~18% at a threshold that identified 20% of the CD cases, but dropped to ~3% at a threshold necessary to identify 85% of the CD cases. In contrast, performance in more clinically relevant settings with higher CD prevalence was substantially better. For instance, the PPV increased from ~18% at 1% prevalence to ~40% at 3% prevalence, and to ~70% at 10% prevalence, with the sensitivity setting at 20% (Figure 5a). Increasing the GRS sensitivity to 60% resulted in a PPV of 40% (at 10% prevalence), and at a sensitivity of 80% the PPV was ~30%. There were some small differences in the AUC between the prevalence levels, on the order of 1—3 percentage points, however, all settings had AUC≥0.86 (Figure 5b). Since sensitivity and specificity are independent of prevalence, these differences are likely due to the small number of cases in the low-prevalence settings and stochastic variations in the data caused by randomly sampling cases from different ethnicities, as each ethnicity showed slightly different predictability of CD in independent validation, together with clinical heterogeneity resulting from different numbers of cryptic cases in the controls of each cohort.

### NON-DISEASE CASES IMPLICATED PER TRUE DISEASE CASE

Another way to quantify the usefulness of predictive models as diagnostic tools is to evaluate the number of subjects without CD that are incorrectly identified as potential CD cases per each true CD diagnosis, for different levels of clinical risk (prevalence). This measure is equivalent to the posterior odds of not having CD given the genotypes (1 – PPV)/PPV, where a lower number is better (fewer incorrect cases implicated per true CD case). Figure 6 shows that at a sensitivity threshold to detect 20% of CD cases, the odds of incorrectly implicating CD were ~7:1 at prevalence of 1%, decreasing to ~1:2 and ~1:5 at a prevalence of 10% and 20%, respectively. Further, at 10% CD prevalence, odds of incorrectly implicating CD lower than 1:1 were achievable with a sensitivity of more than 30%, and for 20% CD prevalence up to 80% of true CD cases could be detected with such odds.

### APPLICATION OF THE GENOMIC RISK SCORE

The diagnostic application of our approach is straightforward: once the SNPs in the model have been genotyped for a given patient, the GRS can be easily computed as the sum of the SNPs weights times the allele dosages plus an intercept term (Methods and Supplementary Table 1). Our models consist of ~200 SNPs, hence the score can be easily computed in a spreadsheet or with PLINK. Whereas the models are fixed in the training phase, the interpretation of the scores depends on the screening setting in which they are used, as selection of different risk thresholds leads to different false positive and false negative rates. In other words, the same numerical risk score may be interpreted differently in each setting, depending on the performance criteria required by the clinician, such as a minimum level of sensitivity or a maximum number of non-CD implicated per true CD implicated.

Figure 7 illustrates how the GRS could be applied in two commonly encountered but different clinical settings, to (i) exclude individuals at average (background) risk of CD with high confidence, or to (ii) stratify individuals at higher risk of CD for further confirmatory testing. In the first setting, in order to optimize the NPV, a suitably low GRS threshold is selected, leading to a relatively large proportion of the population being considered as potentially at-risk of CD. An NPV of 99.6% (comparable to HLA testing) can be achieved at the population-wide 1% prevalence, by setting a threshold corresponding to designating 15% of the population as CD cases (PPV of 5%). In the second setting, we modeled a scenario where the risk of CD is increased (for instance in patients with suggestive symptoms or clinical conditions), and risk stratification to identify the patients most likely to benefit from further definitive investigation for CD is sought. The prevalence of CD in those with higher-risk symptoms is approximately 3% [3,44] and in first-degree relatives of CD patients it is 10% [41,42]. In this second setting, we highlight two extreme choices of threshold, as an example of what is achievable using the GRS at each prevalence level. The first threshold is stringent, predicting only a small number of high-confidence individuals as likely to have CD, and subsequently leading to low sensitivity but to higher PPV. The second threshold is low, implicating a larger number of individuals as likely to have CD, leading to higher sensitivity at the expense of reduced PPV (due to false positives).

More detailed results for a range of prevalence levels (1%, 3%, 10%, and 20%) are shown in Supplementary Table 2. These consider different cutoffs of the risk score, expressed as a proportion of the population considered to be implicated for CD. We used the proportion of the population rather than proportion of the cases (sensitivity) to select risk thresholds, since the true number of cases is unknown and we must select how to classify a given individual based only on their score relative to the population scores estimated in our data (Figure 4). As expected, sensitivity and specificity remain unchanged between the prevalence levels using the same risk score cutoff, however, PPV, NPV, and consequently the number of people incorrectly implicated to have CD for each true CD case, depend strongly both on the prevalence and on the cutoff. Therefore, at a given prevalence level, determined by the population to which the risk score is applied, a suitable risk score cutoff can be selected in order to balance the two competing requirements of increasing the number of people correctly identified as having CD per true cases (PPV), while maintaining an acceptable level of sensitivity (coverage of the cases). A major benefit of the GRS is its flexibility in adapting to the appropriate clinical scenario and needs of the clinician. The PPV of the GRS can be adjusted up or down by varying the GRS cutoff and considering the acceptable level of sensitivity to detect CD. In practice, the most clinically appropriate cutoff thresholds would be determined in local populations by undertaking prospective validation studies utilizing the GRS (see Discussion).

## RISK SCORE CALIBRATION

While the raw GRS cannot strictly be interpreted as the probability of disease given the genotypes, as it is not normalized to be between 0 and 1, the score can be transformed into a probability using the empirical distribution of scores in the data (Figure 4). To assess the agreement between the predicted probability of disease and the observed probability of disease, we used calibration plots [45], comparing the predicted 5% quantiles of the risk scores, derived from models trained on the UK2 dataset and applied to the other datasets (Finn, IT, NL, and UK1), with the observed probability of cases in each bin. For a well-calibrated GRS, the proportion of cases to samples in each bin should be approximately equal to the predicted risk. For example, for a predicted risk of 10%, approximately 10% of the samples in this bin should be cases. To correct for potential lack of calibration, we fitted a LOESS smooth to the calibration curve, which was then used to adjust the raw predictions into calibrated predictions. To avoid biasing the calibration step and to assess how well it performed in independent data, we randomly split each dataset (Finn, IT, NL, and UK1) into two halves of approximately equal size. We assessed calibration in the first half of each dataset, and fitted a LOESS smooth to the calibration curve (Supplementary Figures 4a and 4c). We then used the LOESS smooth to calibrate the predictions for the other half of each dataset, and assessed the calibration there (Supplementary Figures 4b and 4d). Since the calibration is affected by the prevalence, we assessed this procedure both in the observed data (prevalence of ~40%), and in a subsampled version with prevalence of ~10%. Overall, our calibration procedure was able to correct for a substantial amount of mis-calibration in the raw scores, even in the more challenging case of 10% prevalence.

# DISCUSSION

In this study, we have sought to exploit the strong genetic basis for CD and leverage comprehensive genome-wide SNP profiles using statistical learning to improve risk stratification and diagnosis of CD. Our models showed excellent performance in cross-validation (AUC up to 0.90), and were highly replicable in independent replication across datasets of different ethnicities (AUC of 0.86—0.9), suggesting that the genetic component is shared between these European ethnicities and that our models were able to capture a substantial proportion of it. Importantly, even without explaining a majority of CD heritability, the models were robust and accurate, showing that it is not necessary to explain most of the heritability in order to produce a useful model.

The most frequently employed tools to diagnose CD are serology and small bowel histology, but both have limitations. Differences in the sensitivity of antibody recognition of commercially employed CD-specific antigens such as tissue transglutaminase, deamidated gliadin peptides, and endomysial antigen, as well as the human operator performing the assay, can all influence findings and affect reproducibility of serological testing [9,46-49]. Serologic testing in children is reported to be less reliable before the age of 4, and up to 50% of children normalize elevated antibodies over time [50,51]. While small bowel histology remains the 'gold standard' confirmatory test, it is dependent upon patients willing and available to undergo endoscopy, adequate sampling by the gastroenterologist, and appropriate pathological processing and interpretation [52-54]. The frequencies of false positives and false negatives in CD serology assays vary widely and also partly depend upon what degree of histologic inflammation is considered compatible with CD [52,54-58]. Notably, the accuracy of both serologic and histologic testing for CD is dependent on the ongoing consumption of gluten. It is clear that clinically significant variability exists in serologic and histologic work-up for CD and tools to improve the accuracy of CD diagnosis would be of benefit to clinicians. Genomic-based tools are logical given the strong genetic basis for CD and appealing because they are robust and not subject to the kind of variability seen with serologic and histologic assessment, are age-independent, and do not rely on dietary intake of gluten.

A major shortcoming of clinical HLA typing for risk prediction of CD is its poor specificity. HLA testing would result in virtually all CD cases detected, but at the cost of approximately 30—56 people incorrectly implicated for each true case of CD. A significant advantage of the GRS approach is that it can be tuned to the clinical scenario in order to maximize PPV and diagnostic accuracy. By promoting accurate clinical stratification, the GRS could reserve invasive and more expensive confirmatory testing for those who are most likely to benefit from further investigation to secure a diagnosis, and avoid unnecessary procedures in those who are HLA susceptible but unlikely to have CD. This provides both clinical and economic benefits. HLA typing does not provide the flexibility afforded by the GRS, and cannot be effectively employed to identify those who would benefit from endoscopy. For instance, for the sake of comparison, if HLA typing were used as a guide for further investigations, at 10% CD prevalence it would generate over five unnecessary endoscopies per correct endoscopy and at 1% CD prevalence it would generate 30—56 unnecessary endoscopies. Small bowel endoscopy is not a trivial undertaking – the procedure is costly (approximately AUD $750—$1000 for the procedure and associated pathology), has potential complications, necessitates a full day off work, and many patients are reluctant to undergo it.

The GRS can be used to exclude patients unlikely to have CD with a performance comparable to HLA typing (NPV>99% and comparable PPV, Supplementary Table 2). Testing with these parameters may be useful in the clinical scenario of assessing individuals at average risk of CD. A common example would be when a person has commenced a gluten-free diet prior to assessment for CD by serology or small bowel examination and are unwilling or unable to resume oral gluten intake in order to make testing reliable. This is an increasingly common clinical dilemma as the number of people following a gluten-free diet without adequate initial testing for CD continues to rise. In the United States approximately 30% of the adult population are interested in cutting back or avoiding dietary gluten [59].

The GRS can also be used to stratify the risk for CD in patients who present with suggestive clinical features. These risk factors include having a first-degree relative with CD, or problems such as recurrent abdominal pain, bloating, diarrhea or constipation, fatigue, weight loss, unexplained anemia, autoimmune disease (including thyroid disease, T1D, autoimmune hepatitis, rheumatoid arthritis, and Sjogren's syndrome), infertility or early-onset osteoporosis [3,60]. Supporting the recently revised diagnostic guidelines for CD, which promote HLA testing as the 1st line investigation for higher-risk cases, genetic testing of CD is likely to be more informative in these sub-populations exhibiting higher-than-normal prevalence. While clinical guidelines recommend screening for CD in these high-risk populations [61], testing often poses a diagnostic dilemma, as serologic assessment alone cannot confidently exclude a diagnosis, especially given the higher pre-test probability. HLA typing is not particularly informative as the CD HLA susceptibility haplotypes HLA-DQ2.5 and DQ8 are commonly present (manifesting in over 90% of patients with T1D and in 65% in first-degree relatives of individuals with CD) [62,63]. Stratifying these higher-risk patients based on genetic information will allow improved identification of those where small bowel biopsy is likely to be informative. Thus a GRS should reduce the number of unnecessary small bowel biopsies in first-degree relatives who carry HLA susceptibility for CD but do not have it. We have found that our CD models had only moderate predictive ability for T1D, which is consistent with previous findings showing some shared genetics between T1D and CD [38]. Despite the substantial overlap of genetic factors for autoimmune disease, the CD models had negligible predictive ability for Crohn's disease and rheumatoid arthritis. These results indicate that our GRS is specific to CD and less likely to incorrectly identify patients with other autoimmune diseases as having CD, but further work is required to determine whether CD can be as confidently predicted in individuals with T1D as it is in non-T1D populations.

Another major clinical challenge that may benefit from genomic risk prediction is determining the natural history of potential CD (formerly termed 'latent CD') when there is serologic but not histologic evidence of CD, and identifying which patients are more likely to develop overt CD with small bowel inflammation [64]. Current practice is to follow-up all patients with immunologic evidence of gluten intolerance in order to capture those who will eventually develop overt disease. An analogous clinical scenario is that of children with positive CD serology, of whom 50% will fail to develop small bowel changes consistent with CD during follow-up [50,51]. In both clinical situations, it is reasonable to expect that a GRS that takes into account the totality of genetic risk and susceptibility for CD might improve stratification of such patients into low or higher risks of developing overt CD. Of course, environmental factors are important in the development of CD and the extent to which environmental versus genetic factors play in the development of overt CD remains unknown. Long-term follow-up studies of patients with potential CD will be necessary to establish the role of genomic risk prediction in this important subgroup.

Future work will look at optimizing our GRS as a tool to predict CD risk. Validation of our model in real-life practice will be important to confirm the clinical benefit of the GRS in conjunction with serology and/or over HLA typing alone, as well as to what extent other clinical predictors such as sex, age, and family history, can contribute to clinically relevant risk prediction. Future prospective studies will enable direct optimization of the clinical utility (accuracy, practicality, throughput, and cost) afforded by the GRS, for example in comparison or conjunction with CD serology. These studies will also provide a rigorous evidence base for suggested clinical guidelines of GRS usage. Importantly, appropriate GRS cut-off levels to maximize diagnostic accuracy (optimal PPV and NPV for each given clinical scenario and CD prevalence) could be obtained by local prospective validation. Such studies can identify the ultimate clinical role for the GRS: whether it can effectively replace HLA typing, and also whether it is a stand-alone test or one to accompany CD serology. Hadithi et al showed that in patients at high-risk of CD the addition of HLA typing to CD serology had the same performance as either testing strategy alone [65], but the greater precision of the GRS over HLA typing may better complement CD serology. Understanding where the GRS fits in the diagnostic algorithm to optimize precision and cost-effectiveness will be essential, as is the role it might play in the diagnostic work-up of CD in populations with lower levels of clinical risk. Health economic modeling will address the cost-benefits of using the GRS in the diagnosis of CD, taking into account the cheaper cost of GRS over HLA typing, and include the downstream benefits of potentially reducing endoscopies (substantial cost savings and value to patients from reduced discomfort) as well as potential improvements in quality of life from the detection of CD.

It may be that other modeling approaches yield improvements in predictive power, for example non-additive models that consider epistatic interactions between SNPs. Another avenue for improvement is considering each CD subtype separately, recognizing potentially different genetic bases for these conditions. Based on our results, we do not expect improvements from increases in sample size alone, unless there are substantial advances in genotyping technologies.

In summary, this study demonstrates that simultaneous modeling of all SNPs using statistical learning was able to generate genomic risk scores that accurately predict CD to a clinically relevant degree. This was despite the models explaining only a minority of disease heritability. Our GRS better enables clinicians to stratify patients according to their risk of CD compared to HLA typing alone and we predict, more accurately determines those suitable for confirmatory testing in the form of small bowel biopsy. Reserving this invasive, time consuming, and costly procedure for higher-risk cases is likely to improve the accuracy, cost, and public acceptance of testing for CD, and by extension, benefit the overall diagnosis of CD in the community. By better prioritizing higher-risk patients for confirmatory testing, genomic risk prediction carries promise as a clinically useful tool to add to the clinician's diagnostic armamentarium. Ultimately, we envisage a clinical scoring algorithm based on the combination of clinical features, serologic, and genetic information that will accurately predict people with biopsy-confirmed CD, and perhaps ultimately overcome the reliance on small bowel histology altogether. Further, the costs of genotyping a select number of marker SNPs with a low-plex, high throughput technology are already far lower than the costs of full HLA typing, resulting in a test that is cheaper, more flexible, and more precise than HLA typing. More generally, this study demonstrates that statistical learning approaches utilizing SNPs can already produce useful predictive models of a complex human disease using existing genotyping platforms assaying common SNPs, and suggests that similar approaches may yield comparable results in other complex human diseases with strong genetic components.

# Methods

## Ethics Statement
All participants gave informed consent and the study protocols were approved by the relevant institutional or national ethics committees. Details given in references van Heel et al [23] and Dubois et al [24]. All data was analyzed anonymously.

## Data
We analyzed six CD datasets: *UK1* [23], *UK2*, *IT*, *NL*, and *Finn* [24], and *IMM* [66]. The main characteristics of the datasets are listed in Table 1. In addition we used three WTCCC datasets (T1D, Crohn's, and RA) that have been described elsewhere [30,37]. UK1 used the Illumina Hap300v1-1 array for cases and Hap550-2v3 for controls, UK2 used the Illumina 670-QuadCustom-v1 for cases and 1.2M-DuoCustom-v1 for controls, the NL and IT datasets used the Illumina 670-QuadCustom-v1 in both cases and controls, and the Finn dataset used the Illumina 670-QuadCustom-v1 for cases and Illumina 610-Quad for controls. The WTCCC data (T1D, Crohn's, and RA) used the Affymetrix 500K array. In all of our models, we used autosomal SNPs only, and did not include the sex as a covariable, as models built separately on the two genders using the same sample size and case:control balance showed very similar performance in cross-validation on the UK2 dataset (results not shown). For analyses of the MHC region, we defined the MHC as all SNPs on chr6 in the range 29.7Mb—33.3Mb.

## Quality Control
For each of the UK1, UK2, IT, NL, and Finn datasets, we removed non-autosomal SNPs, SNPs with MAF < 1%, with missingness > 1%, and those with deviations from Hardy-Weinberg Equilibrium in controls $P < 5 \times 10^{-6}$. We also removed samples with missingness > 1%. We tested identity-by-descent between samples in UK1 and UK2 and removed one of a pair of samples with $\hat{\pi} \geq 0.05$ (either between the datasets or within the datasets). The QC for the IMM Immunochip data has been previously described [66]; we estimated 5763 Immunochip samples to have $\hat{\pi} \geq 0.125$ (PLINK IBS) with any UK2 sample, and those were removed, leaving 10,304 Immunochip samples in total, with 18,252 SNPs shared with the UK2 dataset (post-QC). The QC for the WTCCC data (T1D, Crohn's, and RA) has been previously described [30,37].

## Assessment of Population Structure Effects
To assess the impact of potential cryptic population structure, we estimated the top 10 principal components (PCs) for the UK2 with EIGENSOFT 4.2 [67], after removal of regions with high LD (see Text S1 for details). The principal components themselves showed almost no predictive ability (AUC=0.52), and models trained on all SNPs accounting for these PCs showed indistinguishable performance from the non-adjusted model, both in cross-validation on the UK2 dataset and in external validation on the Finn, NL, and IT datasets (Supplementary Figure 5), demonstrating that confounding of our UK2 models by population structure was negligible and was not a contributing factor to the high predictive ability.

## STATISTICAL ANALYSIS

### THE PREDICTIVE MODEL

We used L1-penalized support vector machines (SVM) implemented in the tool SparSNP [26] (https://github.com/gabraham/SparSNP) as the classifiers. The L1-penalized SVM is a sparse linear model, that is, many or most of the SNPs will receive zero weight in the model, as determined by the L1 penalty. The use of a sparse model fits with our prior expectation that in autoimmune disease most SNPs will not be associated with disease status. The inherent sparsity of the model obviates the need for subsequent filtering of SNPs by weight, in order to decide which ones show strong evidence of association and which are spurious, as would be required in a non-sparse (L2-penalized) model. In addition, in extensive simulation and in analysis of real genotype data, including the two celiac disease datasets UK1 and UK2, we have previously shown the advantage of L1-penalized SVMs over commonly used approaches such as polygenic scores (sum of the log odds), linear mixed models (GCTA), and unpenalized logistic regression [30]. The advantage of sparse models over standard linear mixed models in predicting autoimmune disease has been recently confirmed in type-1 diabetes as well [68]. We have also shown that our L1-penalized SVMs achieved essentially identical performance to L1-penalized logistic regression (glmnet) in cross-validation over the Finnish subset of the celiac disease dataset, while being substantially faster [26]. Unlike single marker approaches that estimate the effect size of each SNP separately, the L1-penalized SVM is a multivariable model, where the estimated effect of each SNP is conditional on all other SNPs, thereby implicitly accounting for the linkage disequilibrium (LD) between SNPs. Besides imposing sparsity, the L1 penalty tends to produce models where one representative SNP is selected out of a group of highly correlated SNPs, while the rest remain with a zero weight, in contrast with L2-penalized or unpenalized models where many or all of these SNPs may receive a non-zero weight. For an in-depth discussion of these issues and the effects of varying LD levels on the performance of multivariable models, see [30].

The L1-penalized SVM model is induced by minimizing the L1-penalized squared-hinge loss over $N$ samples and $p$ SNPs,

$$L(\beta_0, \beta) = \frac{1}{2N} \sum_{i=1}^{N} \max\{0, 1 - y_i(x_i^T \beta + \beta_0)\}^2 + \lambda \sum_{j=1}^{p} |\beta_j|,$$

where $x_i$ is the $p$-vector of genotypes for the $i$th sample in allele-dosage coding {0, 1, 2}, $y_i$ is the binary phenotype {−1, +1}, $\beta$ is the $p$-vector of weights, $\beta_0$ is the intercept (also called the bias, which is not penalized), and $\lambda$ is the L1 penalty. We also investigated adding an L2 penalty to the model (elastic-net), however, based on initial cross-validation experiments, we found no advantage in the L2 penalty and subsequently did not use it. All of our models were additive in the allele dosage {0, 1, 2}.

The genomic risk score $\hat{y}_i$ for a new sample $x_i$ consisting of $p$ genotypes is then

$$\hat{y}_i = \beta_0 + \sum_{j=1}^{p} x_{ij} \beta_j,$$

where the continuous value $\hat{y}_i$ is later thresholded at different values to produce a binary predicted class. The model was evaluated over a grid of penalties, in 10-fold cross-validation, repeated 10 times. The optimal number of SNPs in the model was decided based on the model with the highest average AUC across the replications. The final model was a consensus model, averaged over all 10x10=100 models, and containing approximately the number of SNPs determined earlier. Post processing and plotting of the results was performed in R [69], together with the packages ggplot2 [70] and ROCR [71].

*MEASURES OF PREDICTIVE PERFORMANCE*

To quantify the predictive performance of the models in cross-validation and external validation, we employed receiver operating characteristic (ROC) curves (sensitivity versus 1 minus specificity), the area under the ROC curve (AUC) [72], and the proportion of phenotypic variance explained [32].

To quantify predictive performance in different population settings, we used the positive and negative predictive values, which can be estimated as

$$PPV = \frac{sens \times prev}{sens \times prev + (1 - spec) \times (1 - prev)},$$

and

$$NPV = \frac{spec \times (1 - prev)}{spec \times (1 - prev) + (1 - sens) \times prev},$$

where "sens" is the sensitivity = TP / (TP + FN), "spec" is the specificity = TN / (FP + TN), and "prev" is the prevalence. The PPV/NPV are equivalent to the posterior probability of a person having /not having the disease given a positive/negative diagnosis, respectively. When the PPV and precision are estimated in data with identical prevalence (that is, the observed prevalence in the data is identical to the prevalence in the population for which we wish to estimate PPV), they are equivalent. Precision is defined as TP / (TP + FP).

# ACKNOWLEDGEMENTS

We thank the investigators of the van Heel et al., 2007, Dubois et al., 2010 and Trynka et al., 2011 papers (David van Heel, Cisca Wijmenga, and Lude Franke) for providing the celiac disease data.

JT-D was supported by an NHMRC Postgraduate Medical Scholarship. AK acknowledges support by National ICT Australia (NICTA). NICTA is funded by the Australian Government through the Department of Communications and the Australian Research Council through the ICT Centre of Excellence Program. MI was supported by an NHMRC early career fellowship 637400. MI and GA were supported by University of Melbourne funding. This research was supported by a Victorian Life Sciences Computation Initiative (VLSCI) grant number VR0126 on its Peak Computing Facility at the University of Melbourne, an initiative of the Victorian Government, Australia.

# TABLES

**Table 1:** List of celiac disease datasets used in this study.

| Name | Ethnicity | Platform | Autosomal SNPs post-QC | Male | Female | Cases | Controls | Total samples post-QC |
|---|---|---|---|---|---|---|---|---|
| **Finn** | Finnish | Illumina | 513,952 | 1206 | 1270 | 647 | 1829 | 2476 |
| **IT** | Italian | Illumina | 515,641 | 332 | 708 | 497 | 543 | 1040 |
| **NL** | Dutch | Illumina | 515,169 | 752 | 897 | 803 | 846 | 1649 |
| **UK1** | British | Illumina | 301,659 | 938 | 1262 | 778 | 1422 | 2200 |
| **UK2** | British | Illumina | 515,444 | 2954 | 3831 | 1849 | 4936 | 6785 |
| **IMM** | British | Immunochip | 18,252* | 3927 | 6377 | 5907 | 4397 | 10,304 |

**\*** only SNPs in common with the post-QC UK2 dataset were analyzed and are thus shown here

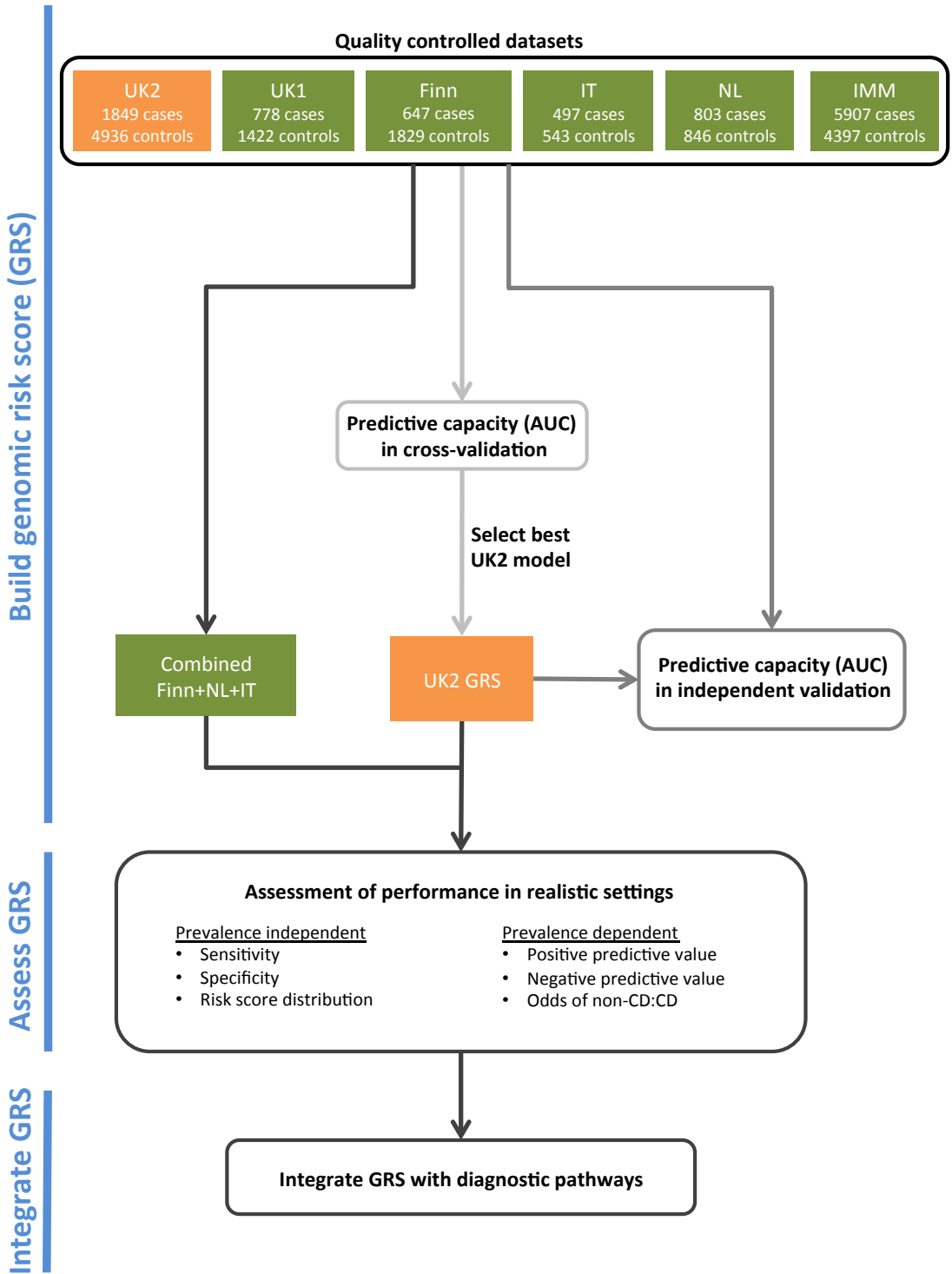

**Figure 1:** The analysis workflow

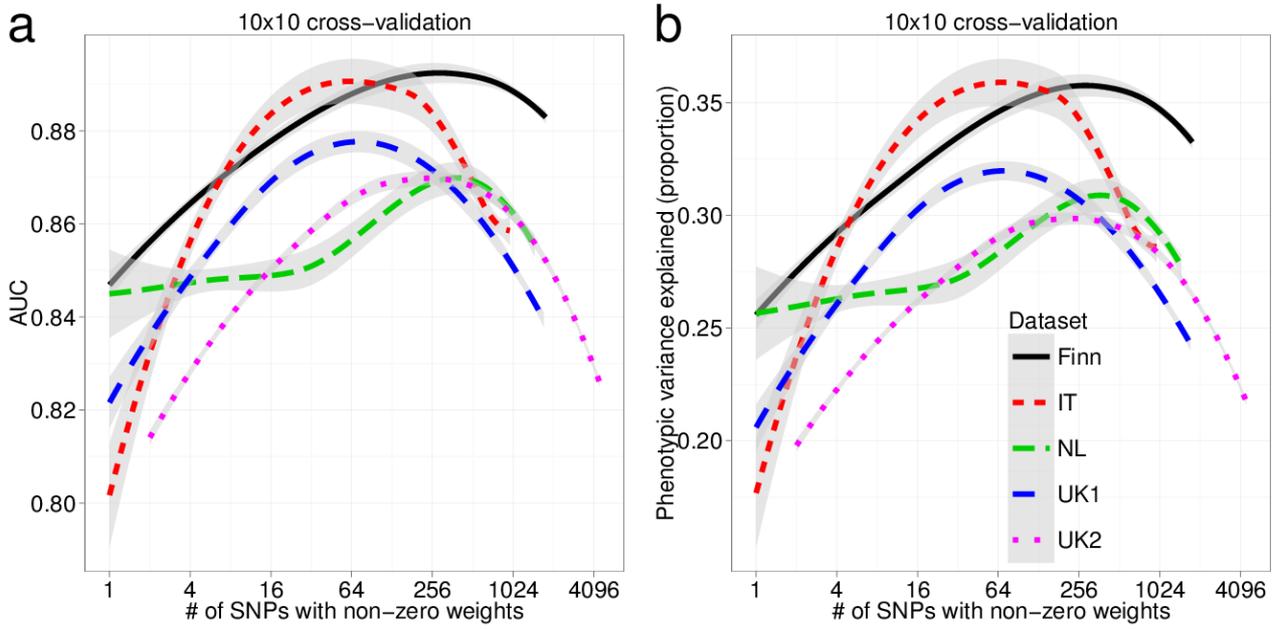

**Figure 2:** LOESS-smoothed (a) AUC and (b) phenotypic variance explained, from 10x10 cross-validation, with differing model sizes, within each celiac dataset. The grey bands represent 95% confidence intervals about the mean LOESS smooth.

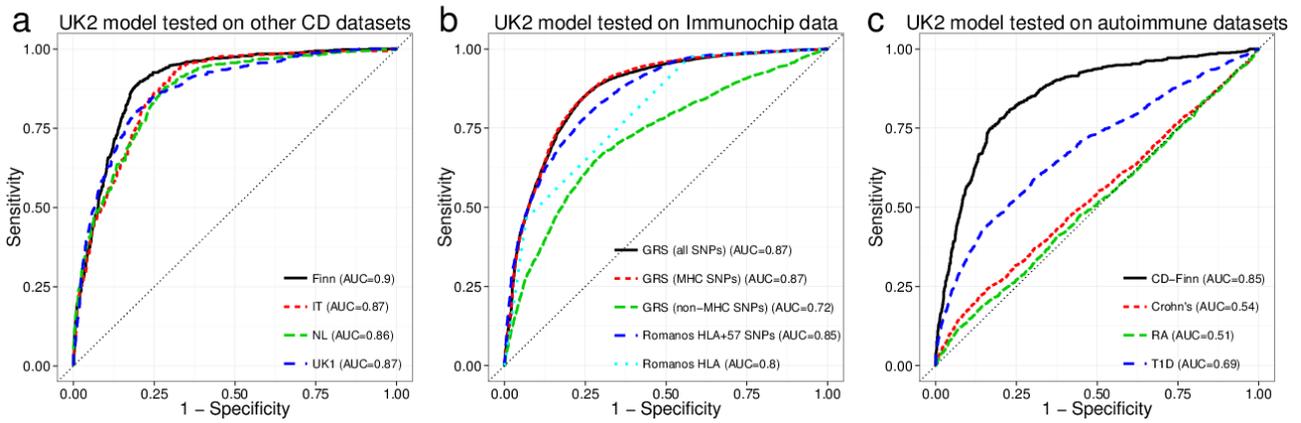

**Figure 3:** ROC curves for models trained in the UK2 dataset and tested on (a) four other CD datasets, (b) the Immunochip CD dataset, comparing the GRS approach with that of Romanos et al. [21], and (c) three other autoimmune diseases (Crohn's disease, Rheumatoid Arthritis, and Type 1 Diabetes). We did not re-tune the models on the test data. For (b) and (c), we used a reduced set of SNPs for training, from the intersection of the UK2 SNPs with the Immunochip or WTCCC SNPs (18,252 SNPs and 76,847 SNPs, respectively). In (c), the same reduced set of SNPs was used for the CD-Finn dataset, in order to maintain the same SNPs across all target datasets.

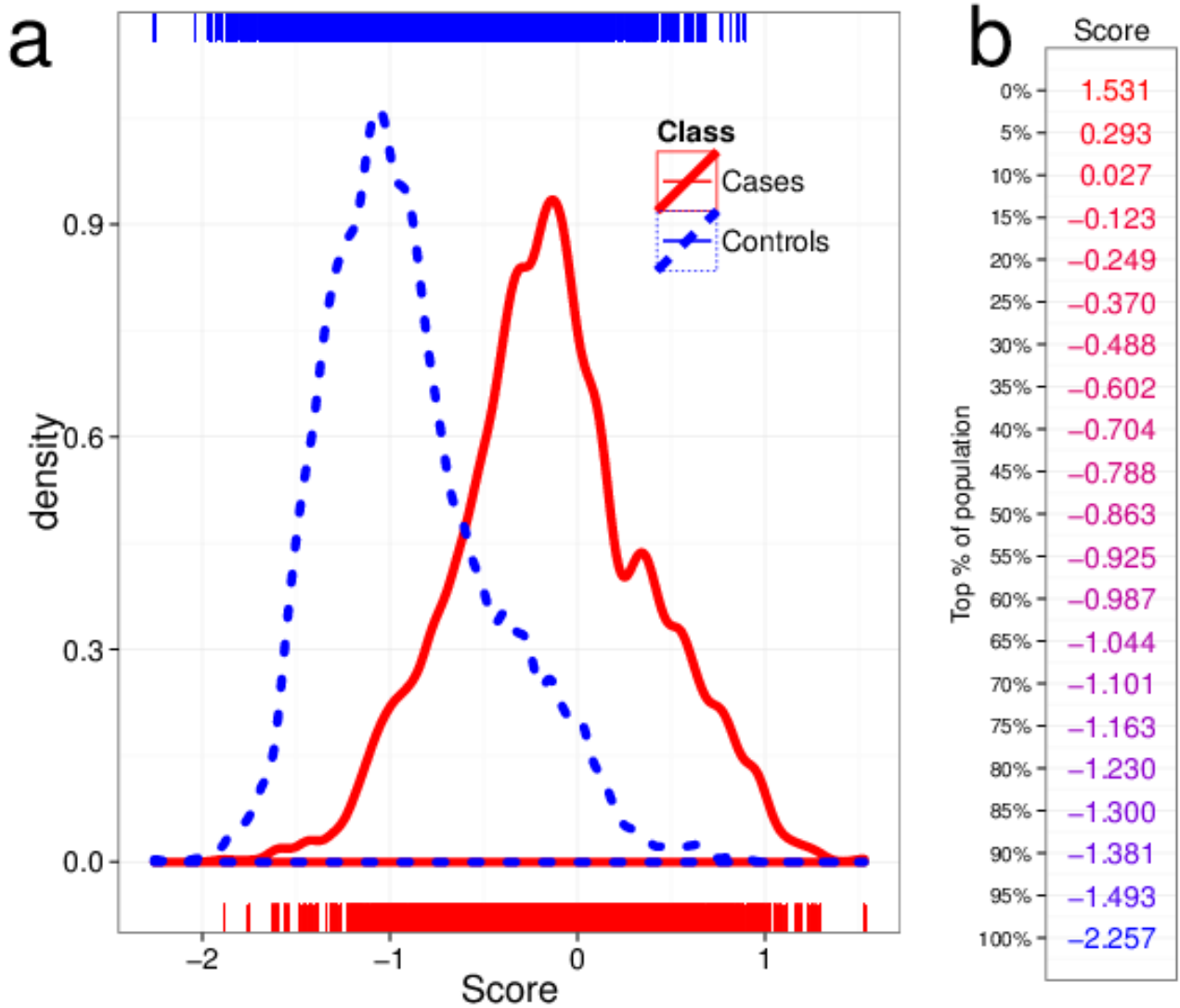

**Figure 4:** (a) Kernel density estimates of the risk scores predicted using models on UK2 and tested in the combined dataset Finn+NL+IT, for cases and controls. (b) Thresholds for risk scores in terms of population percent, with the top more likely to be a CD and the bottom more likely to be non-CD.

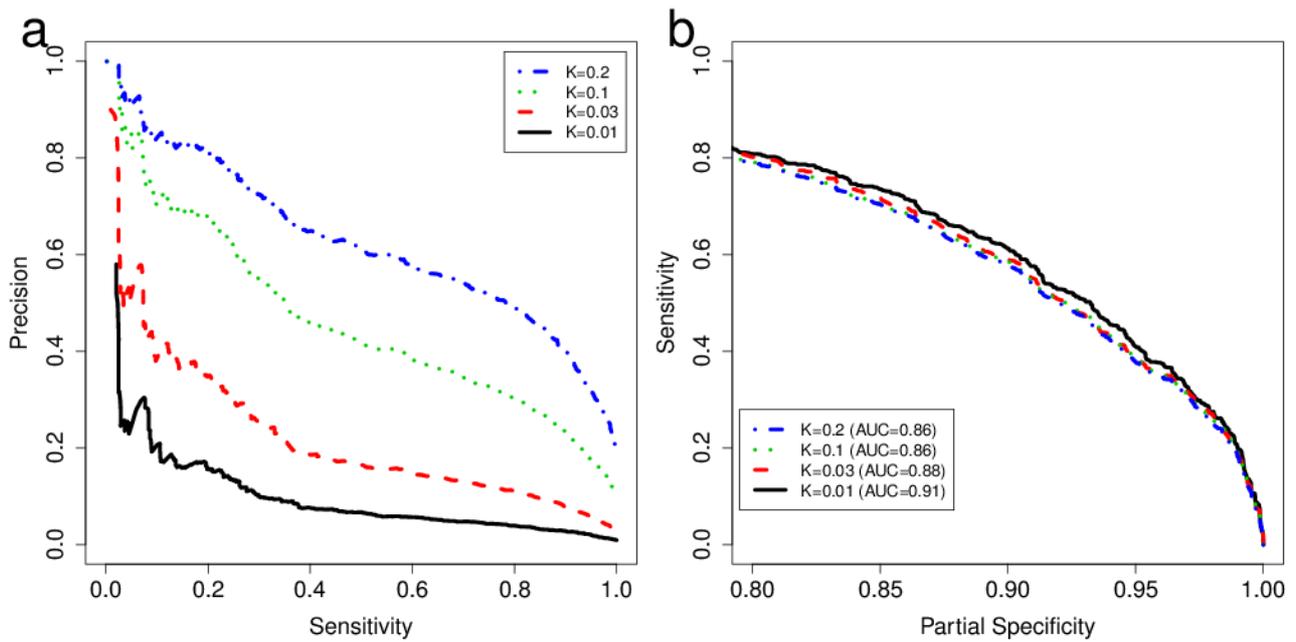

**Figure 5:** (a) Positive and negative predictive values and (b) partial ROC curves for models trained on UK2 using 228 SNPs in the model, and tested on the combined Finn+NL+IT dataset. *K* represents the prevalence of disease in the dataset and the curves are threshold-averaged over 50 replications. Note that precision is not a monotonic function of the risk score. Precision is equivalent to PPV here. A prevalence of ~10% corresponds to prevalence in first-degree relatives of probands with CD [42].

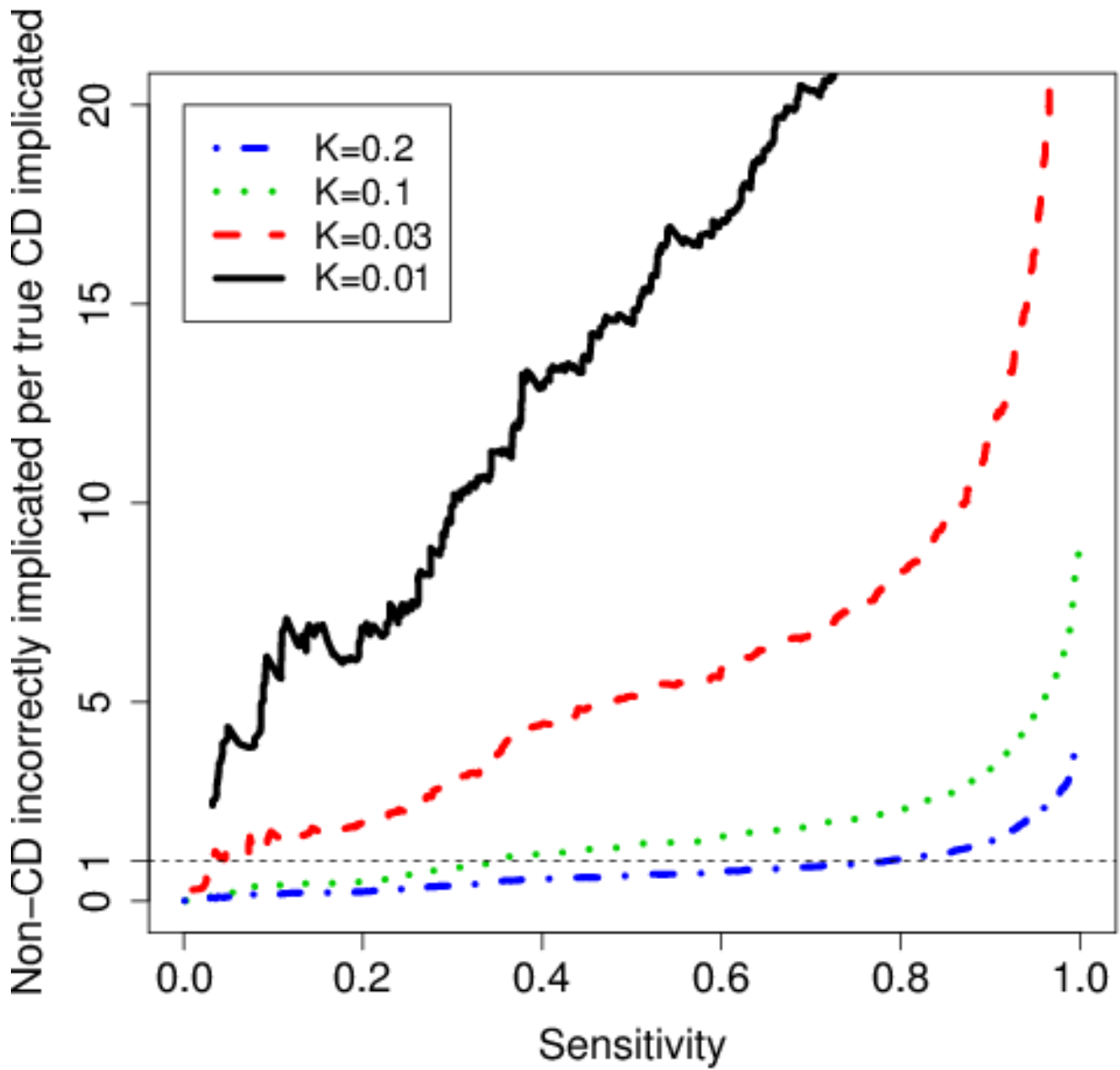

**Figure 6:** The number of non-CD cases "misdiagnosed" (wrongly implicated by GRS) per true CD cases "diagnosed" (correctly implicated by GRS), for different levels of sensitivity. The risk score is based on a model trained on the UK2 dataset, and tested on the combined Finn+NL+IT dataset. The results were threshold-averaged over 50 independent replications. Note that the curve for *K*=1% does not span the entire range due to averaging over a small number of cases in that dataset.

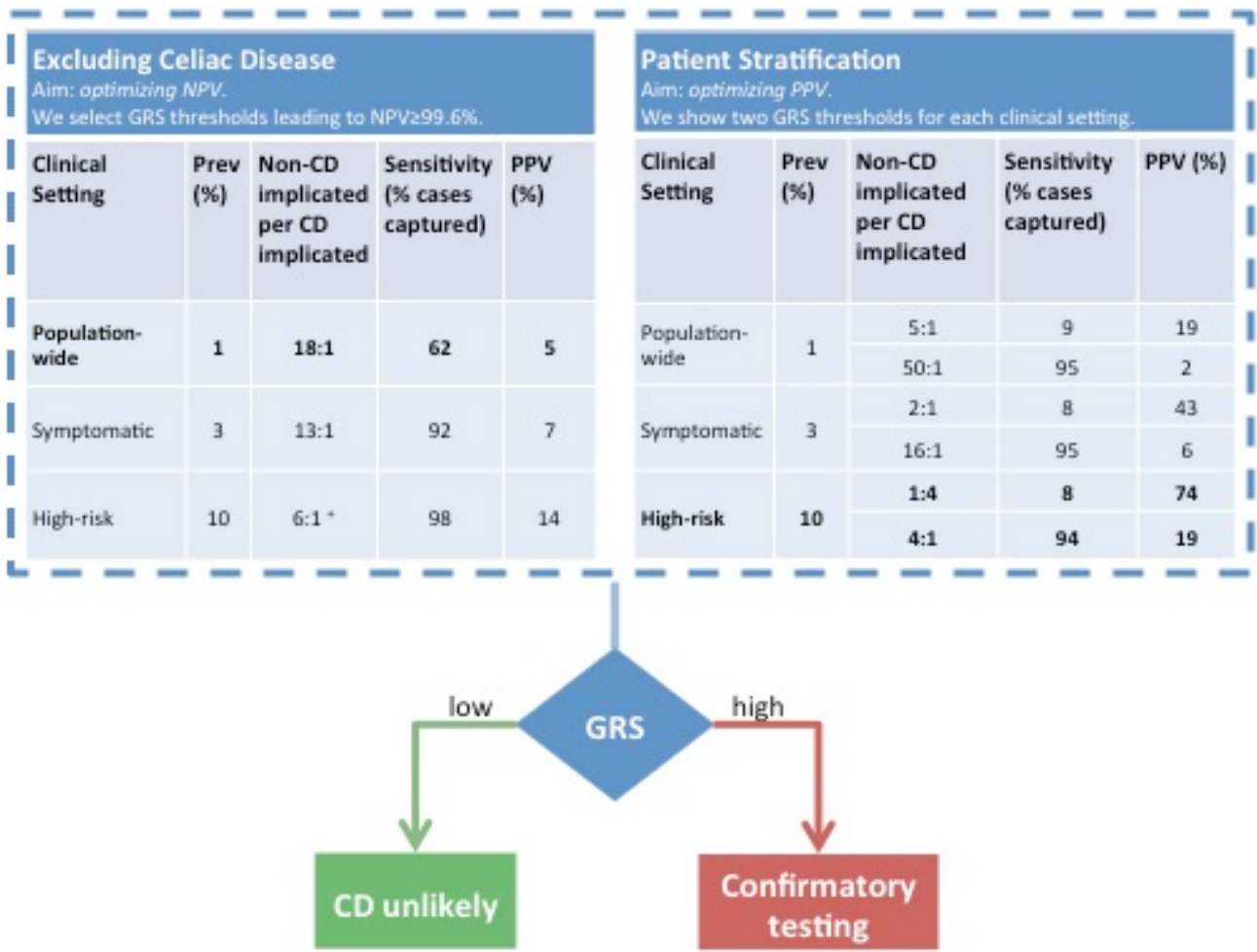

**Figure 7:** The GRS can be employed in different clinical scenarios and tuned to optimize outcomes. The GRS can be employed in a comparable manner to HLA testing (left table) to confidently exclude CD. In this scenario, we selected a GRS threshold based on NPV=99.6% however a range of thresholds can be selected to achieve a high NPV (see note below). The GRS can also stratify CD risk (right table). Confirmatory testing (such as small bowel biopsy) would be reserved for those at high-risk. In this example, we present two scenarios: optimization of PPV or of sensitivity. In comparison to the GRS, all HLA-susceptible patients will need to undergo further confirmatory testing for CD. For more information on GRS performance across a range of thresholds, see Supplementary Table 2. Prospective validation of the GRS in local populations would enable the most appropriate settings for NPV, PPV and sensitivity to be identified which provide the optimal diagnostic outcomes

+ The highest achievable NPV at 10% prevalence was 99.4%.